\documentclass[preprint]{emulateapj}
\usepackage{graphicx}
\usepackage{epstopdf}
\usepackage{epsfig}
\newcommand{\asec}      {\mbox{$^{\prime \prime}  $} }
\newcommand{\amin}      {\mbox{$^{\prime}$}}
\newcommand{\ha}{H$\alpha$}
\newcommand{\hb}{H$\beta$}

\newcommand{\oii}{[O\thinspace{II}]}
\newcommand{\oiii}{[O\thinspace{III}]}

\newcommand{\cii}{C\thinspace{II}}
\newcommand{\ergscm}{$\ erg\ s^{-1}\ cm^{-2}\ $}
\begin{document}
\title{Keck-I MOSFIRE spectroscopy of the $z\sim12$ candidate galaxy UDFj-39546284 $^{\star}$}
 \author{P. Capak\altaffilmark{1}, 
 A. Faisst\altaffilmark{2},
 J. D. Vieira\altaffilmark{3},
 S. Tacchella\altaffilmark{2},
 M. Carollo\altaffilmark{2}, 
 N. Z. Scoville\altaffilmark{3}
}

\altaffiltext{1}{Spitzer Science Center, 314-6 Caltech, Pasadena, CA, 91125}
\altaffiltext{2}{Institute for Astronomy, Swiss Federal Institute of Technology (ETH Zurich), CH-8093 Zurich Switzerland}
\altaffiltext{3}{California Institute of Technology, 314-6 Caltech, Pasadena, CA, 91125}

\altaffiltext{$\star$}{The data presented herein were obtained at the W.M. Keck Observatory, which is operated as a scientific partnership among the California Institute of Technology, the University of California and the National Aeronautics and Space Administration. The Observatory was made possible by the generous financial support of the W.M. Keck Foundation.}

\begin{abstract}
We report the results of deep (4.6h) H band spectroscopy of the well studied $z\sim12$ H-band dropout galaxy candidate UDFj-39546284 with MOSFIRE on Keck-I.  These data reach a sensitivity of $5-10 \times 10^{-19}$\ergscm per 4.4\AA\ resolution element between sky lines.  Previous papers have argued this source could either be a large equivalent width line emitting galaxy at $2<z<3.5$ or a luminous galaxy at $z\sim12$. We find a $2.2\sigma$ peak associated with a line candidate in deep Hubble-Space-Telescope Wide-Field-Camera 3 Infrared grism observations, but at a lower flux than what was expected.  After considering several possibilities we conclude these data can not conclusively confirm or reject the previous line detection, and significantly deeper spectroscopic observations are required. We also search for low-redshift emission lines in ten other $7<z<10$ $z$, $Y$, and $J$-dropout candidates in our mask and find no significant detections. 

\end{abstract}

\keywords{galaxies: high-redshift -- galaxies: formation -- line: identification}

\section{Introduction }\label{s:introduction}

Deep photometric surveys with the Hubble Space Telescope (HST) are revolutionizing our knowledge of the star forming galaxy population  at the epoch of hydrogen re-ionization, only a few hundred million years after the Big Bang.  Hundreds of  candidate $z \sim 7-8$ galaxies have now been discovered using the Lyman break galaxy (LBG or 'dropout') technique using newly available HST Wide-Field-Camera-3-Infrared (WFC3-IR) data \citep{2010ApJ...709L.133B, 2010ApJ...709L..21O, 2013MNRAS.tmp.1319M,2013ApJ...763L...7E}.  Recently, this technique has been pushed to $z\sim12$ by using hundreds of HST orbits to probe to unprecedented depths in the Hubble-Ultra-Deep-Field 2012 (HUDF12) \citep{2013ApJ...763L...7E}.  

The LBG or 'dropout' technique relies on absorption by intervening neutral hydrogen below the Lyman limit at 912\AA\ and Ly$\thinspace \alpha$ at 1216\AA\ to create a strong spectral break that differentiates high and low redshift galaxies using only broad-band photometry.  This technique was first introduced in the 1980's and 1990's \citep{1988prun.proc....1C,1996ApJ...462L..17S}  and broadly adopted as the main technique for finding candidate distant sources once it was shown to be effective spectroscopically on large samples at $z\sim3-4$ \citep{1999ApJ...519....1S,2002ApJ...576..653S} and then deployed at ever higher redshifts \citep{2003PASJ...55..415I,2004ApJ...611..660O, 2007ApJ...670..928B}.  The current frontier is to use the near-infrared Wide Field Camera 3 (WFC3/IR) Hubble Ultra Deep Field data for 'J and H -dropout' galaxies at redshifts $z\sim8-12$ and has led to a handful of tentative detections, most noticeably the source UDFj-39546284, a candidate $z\sim11.9$ galaxy  \citep{2013ApJ...763L...7E} previously claimed to be at $z\sim10.3$ \citep{2011Natur.469..504B,2012ApJ...745..110O}.  If confirmed to be at such high redshifts, even this single galaxy would be of paramount importance to probe the physics of the  earliest phases of  galaxy formation, and to quantify how star forming galaxies  contribute to the re-ionization of the Universe \citep{2010Natur.468...49R,2011Natur.469..504B, 2012ApJ...745..110O}. 

However, spectroscopic redshifts are fundamental to ascertain the true redshifts of candidate high-$z$ LBGs.  At $z>4$ application of the dropout technique has lead to notably diverse results  \citep{2003PASJ...55..415I,2004ApJ...611..660O,2007ApJ...670..928B, 2010A&A...523A..74V} with only limited samples being spectroscopically confirmed \citep{2009ApJ...695.1163V,2010MNRAS.408.1628S, 2012ApJ...760..128M}, and the vast majority of spectroscopy failing to detect anything significant.  For example, \citet{2010MNRAS.408.1628S} and \citet{2009ApJ...695.1163V}, two of the largest spectroscopic samples at $z>4$ to date, confirm less than half of their targeted objects.  This is particularly problematic at $z>6$ where exotic objects can contaminate the dropout selection \citep{2011ApJ...730...68C}. 

A growing body of evidence suggests extreme line emitters and unusual evolved galaxies at $z\sim2$ are a contaminant in $z>7$ LBG selections  \citep{2011ApJ...743..121A,2011ApJ...730...68C,2012MNRAS.425L..19H}.  In published spectroscopic studies of $z>7$ galaxies the vast majority of results are null, with a large fraction of detected objects placed at $z<3$ \citep{2011ApJ...730L..35V,2011ApJ...730...68C,2012MNRAS.427.3055C,2012MNRAS.425L..19H,2012ApJ...744...83O,2012MNRAS.427.3055C,2013MNRAS.430.3314B}.  The null results, combined with the type of contamination is worrying because it is from a poorly understood population of objects and so is difficult to include in the simulations required to quantify LBG selection criteria \citep{2011ApJ...730...68C}.  In the near term this highlights the need for deep spectroscopic studies at $6<z<8$ where current spectrographs can hope to confirm existing high-z candidates and characterize contaminating populations.  In the longer run the James Web Space Telescope (JWST) and Thirty-meter class ground based telescopes will be necessary to obtain the requisite spectroscopic samples at $z>8$ and faint fluxes needed to understand these populations.

UDFj-39546284 has been well studied by many authors and is only detected in the F160W H band filter with HST WFC3-IR, even though the deepest currently possible data exists at both bluer and redder wavelengths.  It was first reported by \citet{2011Natur.469..504B} who originally claimed it was at $z\sim10$.  \citet{2013ApJ...763L...7E} recently improved the depth of the HUDF  F160W and the F105W images by 0.2 and 0.5 magnitudes respectively and added deep F140W imaging which overlaps the blue half of F160W.  This new data (HUDF12) favors UDFj-39546284 being at $z=11.9$, but still allows the possibility it could be a strong line emitter at $2<z<3.5$.  Subsequently, \citet{2013ApJ...765L...2B} reported a $2.7\sigma$ line detection at $15990\pm40\AA$ with a flux of $3.5\pm1.3 \times 10^{-18} $\ergscm based on 40,500 seconds of HST WFC3-IR $R\sim140$ grism spectroscopy.  However this detection required significant modeling of the data to remove overlapping spectra and background artifacts. Subsequent re-analysis of the deeper imaging and spectroscopic data by \citet{2013ApJ...765L..16B} now lead them to the conclusion that the galaxy is more likely at $z\sim2$ than $z\sim10-12$ based on the argument that the galaxy would be unusually luminous if actually at $z\sim12$.

If UDFj-39546284 is at low redshift, the photometric constraints indicate it must be a young blue galaxy with strong line emission that has a line flux of $\sim 3 \times 10^{-18} $\ergscm for a typical dwarf galaxy velocity dispersion \citep{2013ApJ...763L...7E, 2013ApJ...765L...2B, 2013ApJ...765L..16B}.  Furthermore, the line reported by \citet{2013ApJ...765L...2B} falls in a region largely free of sky lines and so can be confirmed by ground based spectroscopy which has significantly higher sensitivity and spectral resolution.

In this paper we present R$\sim3630$ $H$-band spectroscopy of UDFj-39546284 with the MOSFIRE multi-object  infrared spectrograph \citep{2010SPIE.7735E..47M,2012SPIE.8446E..0JM} on the Keck-I telescope, reaching $2-3\times$ the sensitivity of the WFC3 grism spectroscopy between sky lines. This new instrument  saw its first-light in April 2012,  and provides a substantial boost in sensitivity relative to previous facilities for studies of very faint distant galaxies. Its high-multiplexing of up to 46 adjustable slitlets  over a field of view of $6\amin \times 6\amin $ enables the simultaneous acquisition of scores of individual sources.  In addition to UDFj-39546284 we give a summary of the results from other objects on our slit mask in Table \ref{t:results}.

We adopt a cosmological model with $\Omega_\Lambda = 0.7$, $\Omega_M = 0.3$, and $h=0.7$ and  magnitudes in the AB system.

\begin{figure*}
\begin{center}
{\includegraphics[scale=0.25]{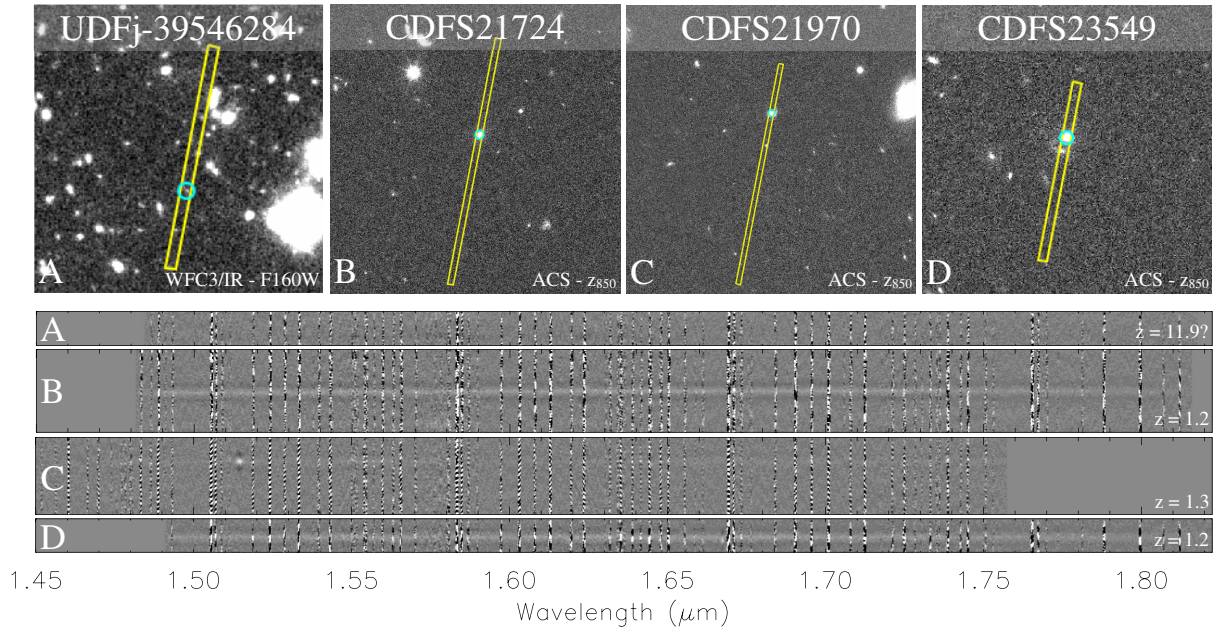}}
\caption{Image cutouts and the 2d MOSFIRE spectra around UDFj-39546284 (A) and three other bright objects are shown.  The MOSFIRE slit positions are marked in yellow and the objects highlighted with a cyan circle.  Analysis of the bright compact-object spectra CDFS21724 (B) indicates we are losing no more than 10\% of the flux due to mask mis-alignment.  The strong line visible in CDFS21970 (C) is \ha\ at $z=1.3089$.  A summary of these and other objects in the mask is given in Table \ref{t:results}.  \label{align}}
\end{center}
\end{figure*}

\section{Data \label{s:data}}
Data were collected on the nights of Jan 15 \&16, 2013 using the MOSFIRE instrument on the Keck-I telescope.  Conditions were photometric on both nights, with a median seeing of 1.2\asec.  The instrument was configured with the H band grating, $0.7\asec$ slit widths, 180s exposures, and 16 Multiple Correlated Double Samples (MCDS).  The telescope was nodded by $\pm1.25$\asec between observations with 44 exposures taken on Jan 15, and 48 on Jan 16, yielding a total exposure time of 4.6h on the mask.  

We used bright 2MASS stars for alignment, but noted a significant $\sim 1\asec$  offset between the 2MASS and HUDF12 astrometry which was corrected before generating the mask.  We verified the alignment stars and galaxies in the HUDF12 were on the same astrometric system by comparing the astrometry of the alignment stars and galaxies on the HUDF12 images and the surrounding HST-ACS images which covered a wider area.  Based on repeated alignment exposures taken every 1-2 hours during both nights the mask was aligned to better than 0.1$\asec$ during the observations.  Finally, to verify the mask was properly aligned, relatively bright objects were placed on slits around the mask and their flux checked against the expectations from photometry (See Figure \ref{align}). 

The data were reduced using the MOSFIRE python reduction package which subtracts nodded pairs of images, then shifts and co-adds the exposures using a sigma-clipped noise weighted mean.  The Argon and Neon arc lamps along with sky lines were used for wavelength calibration.  We generated a combined image of all 92 (4.6h) exposures as well as combinations including only 3/4 of the data (3.45h) to test the robustness of the reductions, noise estimates, and for temporal variations .

\begin{figure}
\begin{center}
{\includegraphics[scale=0.34]{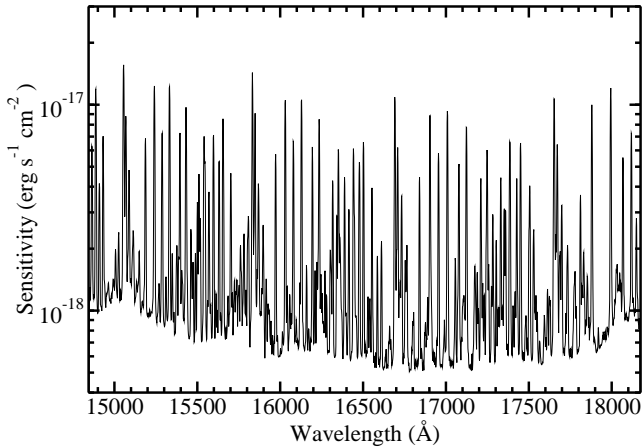}}
\caption{The measured $1\sigma$ sensitivity per 4.4\AA\ resolution element at the position of  UDFj-39546284 is plotted.  The measured sensitivity is consistent with that predicted by the instrument exposure time calculator.  Note that we should be able to detect a $\sim3 \times 10^{-18} $\ergscm line implied by the photometry at several sigma between sky lines. \label{sens}}
\end{center}
\end{figure}

Flux calibration was accomplished by taking spectra of the white dwarf spectrophotometric standard star GD71.  The standard star was observed with a pair of exposures using identical setting to the science observations and reduced in the same way as the science data.  The well detected spectra of the compact, 0.36\asec FWHM in the HST F160W images, $z=1.22$ galaxy CDFS21724 was used to verify our spectrophotometric calibration, slit loss, and noise estimates by comparing this spectra to the ISAAC H band flux from the MUSYC catalog \citep{2010ApJS..189..270C}, the GOODS-S Early Release Science (ERS) F160W flux, and the noise estimates from the MOSFIRE exposure time calculator.  Based on the CDFS21724 spectra additional slit losses due to object extent and mask mis-alignment are estimated to be $5-10\%$ greater than estimated for the standard star.  The noise measured in CDFS21724 between 15920-15950\AA\ is $8.5\times10^{-19} $\ergscm  per 4.4\AA\ resolution element consistent with the estimate of $8.1\times10^{-19} $\ergscm from the exposure time calculator once the slit losses of $48\%$ due to poor seeing, and $10\%$ due to object extension and mask alignment are taken into account.  Two other bright objects CDFS21970 and CDFS23549 were also placed on the mask to verify the throughput and resulted in the expected signal-to-noise, but were not used for quantitative analysis because of a strong emission line (CDFS21970) and complex morphology (CDFS23549) which made a comparison to UDFj-39546284 more difficult.  The estimated sensitivity per resolution element is plotted in Figure \ref{sens} and results for the alignment galaxies and high-z spectra are given in Table \ref{t:results}.

\begin{figure}
\begin{center}
{\includegraphics[scale=0.84]{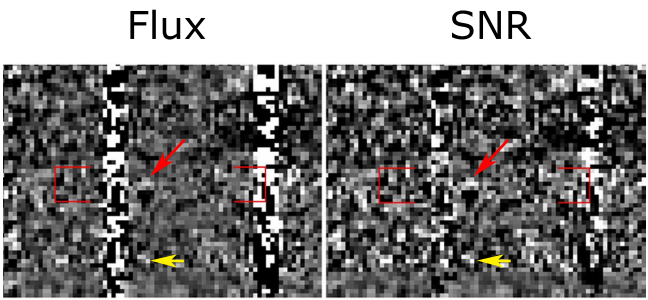}}
\caption{A flux and Signal-to-Noise (SNR) map of the region around the potential detection reported by \citet{2013ApJ...765L...2B}. The region allowed by \citet{2013ApJ...765L...2B} is marked by the red brackets, and the potential $2.2\sigma$ detection in our data is marked by a red arrow.  The region used to subtract the sky from the negative spot below the detection is indicated with a yellow arrow, note the positive fluctuation at this position. \label{det}}
\end{center}
\end{figure}

\begin{figure*}
\begin{center}
{\includegraphics[scale=0.34]{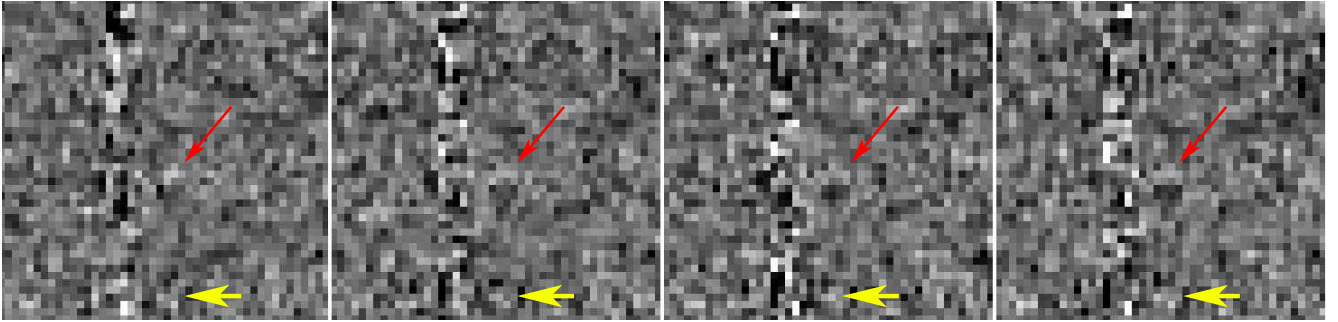}}
\caption{Signal-to-noise maps of the area around the possible line detection with a quarter of the data removed in each reduction.  Note the possible line detection shown in Figure \ref{det} is present in all 4 reductions with a significance of 1.3-2.6 $\sigma$ consistent with the expected noise.  In contrast, the negative spot below the detection varies significantly, indicating it is due to noise.  The regions used to subtract the sky from the negative spot below the detection is indicated with a yellow arrow. \label{jack}}
\end{center}
\end{figure*}

\begin{deluxetable*}{lllll}
\tabletypesize{\scriptsize}
\tablecaption{Targeted objects\label{t:results}}
\tablehead{
\colhead{ID} & \colhead{RA} & \colhead{DEC} & \colhead{$H_{AB}^{d}$} & \colhead{Comments}}
\startdata
CDFS21724$^{b}$		&	$03^h32^m35.636^s$	& $-27^d43^m10.16^s$	& 20.49	&  low-z, Continuum SNR$=9.6$\\
CDFS21970$^{b}$		&	$03^h32^m35.972^s$	& $-27^d48^m50.40^s$	& 21.04	& low-z, Continuum SNR$=1.75$, strong H$_{\alpha}$ line\\
CDFS23549$^{b}$		&	$03^h32^m38.107^s$	& $-27^d44^m32.59^s$	& 20.41	&  low-z, Continuum SNR$=9.2$\\
UDFj-39546284$^{a}$	&	$03^h32^m39.54^s$	& $-27^d46^m28.4^s$	& 29.3	& $z\sim 12$ J drop, no detection\\
UDF12-4106-7304$^{a}$	&	$03^h32^m41.06^s$	& $-27^d47^m30.4^s$	& 29.7	& $z\sim 10$ J drop, no detection\\	
UDF12-3947-8076$^{a}$	&	$03^h32^m39.47^s$	& $-27^d48^m07.6^s$	& 29.0	& $z\sim 10$ J drop, no detection\\
UDFj-43696407$^{a}$	&	$03^h32^m43.69^s$	& $-27^d46^m40.7^s$	& 29.5	& $z\sim 10$ J drop, no detection\\	
UDFj-35427336$^{a}$	&	$03^h32^m35.42^s$	& $-27^d47^m33.6^s$	& 29.6	& $z\sim 10$ J drop, no detection\\	
UDFy-33436598$^{a}$	&      $03^h32^m33.43^s$	& $-27^d46^m59.8^s$	& 29.4	& $z\sim 8$ Y drop, no detection\\	
UDF12-3858-6150$^{c}$	&      $03^h32^m38.58^s$	& $-27^d46^m15.0^s$	& 29.9	& $z\sim 8$ Y drop, no detection\\	
UDF12-3939-7040$^{c}$	&      $03^h32^m39.39^s$	& $-27^d47^m04.0^s$	& 28.9	& $z\sim 8$ Y drop, no detection\\	
UDF12-4057-6436$^{c}$	&      $03^h32^m40.57^s$	& $-27^d46^m43.6^s$	& 28.7	& $z\sim 7$ z drop, no detection\\	
UDF12-3817-7327$^{c}$	&      $03^h32^m38.17^s$	& $-27^d47^m32.7^s$	& 30.3	& $z\sim 7$ z drop, no detection\\	
UDF12-3853-7519$^{c}$	&      $03^h32^m38.53^s$	& $-27^d47^m51.9^s$	& 29.7	& $z\sim 7$ z drop, no detection\\	
\enddata
\tablenotetext{a}{\citet{2013ApJ...763L...7E}}
\tablenotetext{b}{\citet{2010ApJS..189..270C} }
\tablenotetext{c}{\citet{2011ApJ...737...90B,2013ApJ...768..196S}}
\tablenotetext{d}{WFC3-IR F160W for all source names starting with UDF, and ISAAC H for those starting with CDFS.}
\end{deluxetable*}

To find potential emission lines we used SExtractor to automatically search for groups of 9 pixels above $1\sigma$ with no smoothing yielding a $>3\sigma$ net detection, as well as visually inspecting the spectra at the expected position of UDFj-39546284.  We find no robust line detections, but do find a $2.2\sigma$ peak at $15985.5\pm4.4\AA$ with a flux of $1.4\pm0.6 \times10^{-18} $\ergscm in a 6-pixel diameter aperture (Figure \ref{det}), consistent at the1.5$\sigma$ level with the wavelength and flux reported in \citet{2013ApJ...765L...2B}.  We also measured the line flux using an optimal extraction assuming a 1.2\asec gaussian FWHM and found the same result.  This corresponds to $70\pm30\%$ of the measured broad band flux, and would correspond to an observed frame equivalent width of $\sim6400$\AA.

The peak is present in all of the reductions which include only 3/4 of the data (at reduced significance, see Figure \ref{jack}) and measurement of the flux at the positions of the expected negative images due to the dithering return a consistent negative flux, increasing the reliability of the detection.  No known detector artifacts fall on this part of the detector, and no decaying latent images from alignment or previous observations are visible in the early frames of each observation or in the un-dithered stack of the data.  A $1.5\sigma$ negative "spot" corresponding to a region of known bad pixels at one dither position is observed just below the line.  The bad pixels have been masked, but as a result the data at this position only comes from one dither position.  This negative spot is likely a sky-subtraction artifact due to a corresponding positive spot visible at the other dither position where the sky was determined.   This is not an astrophysical object because the expected dual negative, single positive pattern is not observed.  Furthermore, this region varies in intensity in the 3/4 reductions, indicating it is either a noise fluctuation, or a transient effect in the detector.  The region does not produce the adjacent positive detection observed for UDFj-39546284 in other masks taken on the same nights.

 Despite the positive evidence, this detection should be considered tenuous  because 291 other $>2\sigma$ peaks are found between sky lines, 37 of which occur in the full stack and all four 3/4 time stacks, and 4 of which have two negative detections at the two expected dither positions.  This places the probability of a chance co-incidence between a noise peak and the reported WFC3 grism observations at between $1\%$ and $86\%$ depending on wether we assume there are 4 or 291 $>2\sigma$ peaks.   In addition the peak is near a sky line increasing the chance of residuals.
    
Besides UDFj-39546284, we observed 4 J dropout, 3 Y dropout, and 3 z dropout galaxies taken from \citet{2013ApJ...763L...7E} and \citet{2013ApJ...768..196S} with details noted in Table \ref{t:results}.  The goal of including these sources was to search for strong emission lines if the sources were actually at $z<3$ since all are expected to have Ly-$\alpha$ blue ward of our spectroscopic range.  We find no lines at a sensitivity within 10\% of that indicated in Figure \ref{sens} placing useful limits for future spectroscopic observations seeking Ly-$\alpha$ in the Y and J bands. 

  \section{Discussion \label{s:disc}}

Considering the possibility that UDFj-39546284 is a $z<<10$ galaxy, the HST observations imply a line flux of $3-4.7 \times 10^{-18} $\ergscm.   If we combine our 2.2$\sigma$ flux measurement with the 2.7$\sigma$ result of \citet{2013ApJ...765L...2B} it results in a $3.5\sigma$ detection at $15985.5\pm4.4\AA$ with a flux of $1.8\pm0.5 \times10^{-18} $\ergscm, at the lower-end of what is required to explain the photometry.  If we take this combined result at face value along with the photometric analysis of \citet{2013ApJ...765L..16B} it indicates UDFj-39546284 is likely at $z\sim2.19$, and we are seeing the \oiii\ 5007\AA\ line.  But, at the low significance of our data and that of \citet{2013ApJ...765L...2B} confirmation bias due to co-incident noise peaks is problematic and significant unknown systematic effects could be present in both reductions.  Furthermore, for typical \oiii/\hb\ line ratios of $7.3$ in strong line emitting galaxies \citep{2007ApJ...668..853K} one would expect a $>2\sigma$ detection in F140W filter imaging if this source were at $z\sim2.19$ and this is not observed.  In contrast to \citet{2013ApJ...765L...2B} we find even the extreme object they find with a \oiii/\hb\ line ratio of $11.4$ should also result in a $\sim2\sigma$ detection assuming the \citet{2013ApJ...763L...7E} photometric limits.  The discrepancy is due either to differences in the depths quoted between \citet{2013ApJ...765L...2B} and \citet{2013ApJ...763L...7E} which differ by $\sim0.4$ magnitudes and/or to how the $z=1.606$ extreme object photometry and spectra were compared to the UDFj-39546284 limits.  In particular, we note the spectral energy distribution shown in \citet{2013ApJ...765L...2B} is not directly comparable to the plotted UDFj-39546284 limits because the filters in the two observations have significantly different bandwidths and the spectral energy distribution is dominated by lines which create significant non-linear flux changes as they redshift through the filter bandpasses.   An alternative low redshift explanation is that we are detecting one of the \oii\ 3727\AA\ doublet lines at $z=3.29$, however this is less likely since the \oii\ line is typically significantly weaker than the \oiii\ 5007\AA\ line.  So more data are required before coming to any firm conclusions.  

We estimate $\sim20h$ of integration in good conditions with Keck-I-MOSFIRE or an equivalent instrument would confirm the tentative detection at 15985.5\AA\ at $>5\sigma$, but still be insufficient to detect other lines for typical \oiii/\hb\ line ratios in high-equivalent width line emitting galaxies \citep{2007ApJ...668..853K}.  If the preliminary detection is the strong \oiii\ 5006.8\AA\ line at $z=2.19$ the 4958.9\AA\ line would be behind a bright sky line and \hb\ would be too faint to detect.  If instead we are seeing the less likely 3727\AA\ \oii\ doublet at $z=3.29$ the second line of the doublet would be hidden behind the nearby sky line. 

One could also integrate for a similar amount of time in the $K$ band, and attempt to detect the \ha\ line if the source is at $z=2.19$ or both of the \oiii\ lines if the source is at $z=3.29$.  Assuming our preliminary detection is real, all three of these lines fall between sky lines and should be detectable at $>5\sigma$ for typical line ratios.  This may be a more productive approach since it can differentiate between $z=2.19$, 3.29, and 12.14.

Spectroscopic confirmation with ALMA would be expensive.  Assuming this source is at $z\sim12$, is un-obscured, and all of the continuum flux is due to star formation, the expected \cii\ 158$\mu$m luminosity would be $\sim50\mu$Jy assuming 1\% of the bolometric luminosity is emerging in this line.  With full ALMA this would require $\sim72$ hours of integration in one receiver tuning to detect at $5\sigma$.  However, a non-detection would not confirm a low-redshift because of the intrinsic scatter in bolometric luminosity to \cii\ line ratio.

If UDFj-39546284 is at high redshift we expect no detection in the longer spectra and it is unlikely that any conclusive spectra can be obtained with current instrumentation.  The typical equivalent width of Ly$\thinspace\alpha$ in $4<z<6$ galaxies is $\sim 20\AA$ with a tail out to $\sim80\AA$  and is expected to decrease at $z\sim6-12$ due to the increasing opacity of the inter-galactic medium (IGM) \citep{2011ApJ...728L...2S,2012ApJ...760..128M}.   Assuming UDFj-39546284 is at $z\sim12$, our flux limit would require a Ly-$\alpha$ equivalent width of $\ge 200\AA$ to yield a $2.2\sigma$ detection.  This large an equivalent width is occasionally observed in the $z\sim4-6$ universe, but less likely at $z\sim12$ where the inter-galactic medium is expected to be more absorptive.  For a normal equivalent width of $\sim20\AA$, $\sim460$h of observation with Keck-I-MOSFIRE would be required to detect this galaxy, and even an adaptive optics assisted thirty-meter class telescope would require $\sim16$h of integration.

\section{Conclusions \label{s:conclusions}}
Using the Keck-I MOSFIRE infrared spectrograph we recover the $2.7\sigma$ line detection reported in \citet{2013ApJ...765L...2B} at the $2.2\sigma$ level.  Combining these results we find 3.5$\sigma$ line a $15985.5\pm4.4\AA$ with an estimated flux of $1.8\pm0.5 \times10^{-18} $\ergscm, but the likelihood that this is a chance coincidence of noise peaks is non-negligible, so further observations are needed.  If confirmed, the detection indicates UDFj-39546284 is actually a low redshift line emitter at $z\sim2.19$ or 3.29, so both $H$ and $K$ band spectra should be obtained to uniquely determine its redshift.  If the source is at $z\sim12$ it is unlikely ground based 8-10m telescopes or ALMA could confirm it and additional data will yield a non-detection.   The difficulty in interpreting this source, and the high likelihood that it is actually at low-redshift highlights the need for spectroscopic studies at $6<z<8$ to understand possible sources of contamination at even higher redshifts. 

\acknowledgments

We would like to acknowledge the support of the Keck Observatory staff who made these observations possible and the valuable feedback from Janice Lee and Masami Ouchi while preparing this draft.  We also thank Nick Konidaris for providing and supporting the MOSFIRE reduction pipeline and Gwen Rude for providing the MOSFIRE exposure time calculator. AF and ST acknowledge support from the Swiss National Science Foundation; AF also thanks Caltech for hospitality while this article was worked on.  The authors wish to recognize and acknowledge the very significant cultural role and reverence that the summit of Mauna Kea has always had within the indigenous Hawaiian community.  We are most fortunate to have the opportunity to conduct observations from this mountain.

{\it Facilities:}  \facility{Keck:I (MOSFIRE)}

\end{document}